\documentclass[twocolumn,floats,,amsmath,amssymb]{revtex4}
\newcommand{\be}{\begin{equation}}
\newcommand{\ee}{\end{equation}}
\newcommand{\bea}{\begin{eqnarray}}
\newcommand{\eea}{\end{eqnarray}}

\usepackage{graphicx}
\usepackage{dcolumn}

\begin{document}
\draft

\title{An Anderson Impurity Model for Efficient Sampling of Adiabatic Potential Energy Surfaces of Transition Metal Complexes}

\author{M.X. LaBute$^{1,3}$, R.G. Endres$^{2,3}$, and D.L. Cox$^{3}$}

\address{
Theoretical Division, Los Alamos National Laboratory, Los Alamos, NM 87545, USA$^{1}$ \\
Center for Computational Sciences \& Computer Science and Mathematics Division, \\
Oak Ridge National Laboratory, Oak Ridge TN 37831-6164, USA$^{2}$ \\
Department of Physics, University of California, Davis, CA 95616, USA$^{3}$}

\date{\today}

\begin{abstract}
We present a model intended for rapid sampling of ground and excited state potential energy surfaces for
first-row transition metal active sites. The method is computationally inexpensive and is suited for dynamics simulations
where (1) adiabatic states are required "on-the-fly" and (2) the primary source of the electronic coupling between the diabatic states is the perturbative
spin-orbit interaction among the 3$d$ electrons. The model Hamiltonian we develop is a variant of the Anderson impurity
model and achieves efficiency through a physically motivated basis set reduction based on the large value of the $d$-$d$ Coulomb interaction
$U_{d}$ and a L$\acute{a}$nczos matrix diagonalization routine to solve for eigenvalues. The model parameters are constrained by fits to 
the partial density of states (PDOS) obtained from {\it ab initio} density functional theory calculations. For a particular application
of our model we focus on electron-transfer occuring between cobalt ions solvated by ammonium, incorporating configuration interaction
between multiplet states for both metal ions. We demonstrate the capability of the method to efficiently calculate adiabatic potential 
energy surfaces and the electronic coupling factor we have calculated compares well to previous calculations and experiment. 
\end{abstract}

\maketitle

\section{Introduction}
First-row transition metals (TM) complexed with organic ligands play key roles in biology and aqueous chemistry. Ligand exchange and electron transfer
are potentially rate-limiting processes that must be considered in industrial or bio-remedial chemistry \cite{aqueouschemistry}. Ligand binding at the iron porphyrin active
sites of heme proteins are fundamental for oxygen transport and aerobic metabolism \cite{hemes}. Many of these reactions involve nonradiative transitions between
different spin states and require spin-orbit coupling. This interaction occurs primarily on the TM atom due to the localized nature of the 3$d$ valence
electrons and most quantum chemistry codes do not incorporate the small ($\sim$0.05-0.1 eV) energy-scale coupling. \\
\indent
In general, treatment of open-shell TM atoms is difficult and many of the standard  
density-functional theory (DFT) model chemistries such as the local density or gradient-corrected approximations (LDA and GGA, respectively) are known to give erroneous
qualitative results in some cases \cite{ggawrong}. The primary source
of these problems are {\it non-dynamical correlations} associated with the near-degeneracy of configurations with
the Hartree-Fock single-Slater determinant groundstate \cite{yguo}. In fact, it is clear that in the vicinity of the transition state of a
charge transfer complex such {\it non-dynamical correlations} may be very important, much as they are in
chemical bond breaking.  One may incorporate these effects by higher-order corrections such
as Moller-Plesset perturbation theory (MP2,MP4) or configuration interaction (CI) methods such as complete active space self-consistent field (CASSCF), 
or, especially for non-dynamical correlations, multireference configuration interaction approaches\cite{multirefbook},
but these methods are computationally prohibitive for TM molecules containing $\sim$50 atoms. In many cases the CI must be limited to single excitations,
which may be insufficient\cite{quantchem}.

\indent
While the addition of some amount of non-local exchange used in the hybrid Hartree-Fock/DFT methods such as B3LYP seems to capture some
of the multi-determinantal character of TM systems \cite{martinNiO,rosso,rulivsek}, the minimal local Gaussian basis sets required for quantitative accuracy are very large
and the sampling required for dynamics and the calculation of excited states is not yet affordable for most processors. 

\indent
In this paper we introduce an effective semi-empirical model Hamiltonian that is well-suited for ($\sim$50-100 atoms) TM molecules. 
The problematic aspect of strongly correlated TM systems, the ($d$-$d$ Coulomb interaction), leads to a physically-motivated reduction 
of the many-body basis set required by limiting the number of $d^{n}$ configurations that must be considered. We develop a variant of 
the Anderson Impurity Model (AIM), which has had some success in describing spectra and energetics of TM solid-state systems \cite{sawatzky}. We then
describe our method of solving for the groundstate of this model using the variational {\it ansatz} for the groundstate wavefunction
of Gunnarsson and Sch\"{o}nhammer \cite{gandsno1} and Varma and Yafet \cite{varma}. The AIM reduces the multiple-degrees of 
freedom of this problem to a small number of one-electron and Coulomb input parameters. We show how these parameters can be constrained and, in some
cases, directly extracted from {\it ab initio} DFT calculations. We have done this using the SIESTA \cite{siesta} code. The tensor product basis formed 
from the TM 3$d$ and ligand orbitals results in a very large eigenvalue problem. We solve for ground and excited states with an efficient L$\acute{a}$nczos diagonalization
routine. This model, which is similar to the angular-overlap model of Gerloch and co-workers \cite{gerloch}, differs in both the physically-motivated truncation
of the basis set and also in the method of parameterization. We seek here to identify the {\it microscopic} origin of model parameters that describe
the electronic structure, such as the $d$-$d$ Coulomb repulsion and the ligand-field splitting, rather than focusing on structural sensitivities \cite{friesner}

We now develop the model in detail
and apply it to the electron transfer reaction between $Co(NH_{3})_{6(aq)}^{2+}$ [Co(II)] and $Co(NH_{3})_{6(aq)}^{3+}$ [Co(III)] molecules in solution, which
has been extensively investigated by {\it ab initio} electronic structure methods \cite{endreslabute,larsson86,newton1st}.

\section{\label{section2}The Model}
\indent

The TM molecules we wish to investigate are typically composed of a single 3$d$ TM atom bound to one or more ligands. We express
the model Hamiltonian in 2nd-quantized notation using the 3$d$ atomic orbitals of the TM atom and the molecular orbital eigenstates of the ligands.
We have recently applied a similar model to the study of the electronic structure of cobalt valence tautomers \cite{labute}. 

\indent
The electron correlation effects that we wish to incorporate into our model result from the localized character
of the atomic 3$d$-orbitals on the TM atom. The ligand molecular orbitals will tend to be more
delocalized over several atomic sites. A set of localized atomic levels hybridized to a reservoir of more
extended electronic states is well-described by an Anderson Impurity Model. We may write down a variant
of this model, specific for TM molecules: 

\begin{eqnarray}
H &=& \epsilon_{L}\sum_{j\sigma}c_{j\sigma}^{\dag}c_{j\sigma} +
 \sum_{\gamma\sigma}[\epsilon_{d\gamma\sigma} 
 -g_{\gamma}x]d_{\gamma\sigma}^{\dag}d_{\gamma\sigma} \\ \nonumber 
  &+&\sum_{\stackrel{\gamma,i}{\sigma}}V_{\gamma i}(c_{i\sigma}^{\dag}d_{\gamma\sigma} + h.c.)
 +U_{d}\sum_{\gamma\sigma>\gamma^{\prime}\sigma^{\prime}}
 n_{\gamma\sigma}n_{\gamma^{\prime}\sigma^{\prime}} \\ \nonumber
  &+&J_{d}\sum_{\gamma\sigma>
 \gamma^{\prime}\sigma^{\prime}}\vec{s}_{\gamma\sigma}\cdot\vec{s}_{\gamma^{\prime}
 \sigma^{\prime}}
 +\sum_{i=1}^{N_{d}}\xi(\vec{r})\vec{l_{i}}\cdot\vec{s_{i}} \\ \nonumber
  &+&\frac{1}{2}\sum_{\stackrel{i\neq j}{\sigma}}V_{ij}(c_{i\sigma}^{\dag}
 c_{j\sigma} + h.c.) + \frac{1}{2}Kx^{2}\label{H}
\end{eqnarray}

where $\epsilon_{Li}$ are the on-site energies for the redox-active ligand molecular orbitals and $c_{j\sigma}^{\dag}$ are the 
corresponding fermion creation operators for the ligand molecular orbitals. The operator $d_{\gamma\sigma}^{\dag}$ creates an electron
in the TM 3$d$-orbital with $O_{h}$ symmetry label $\gamma = xy, yz, xz, 3z^{2}-r^{2}, x^{2}-y^{2}$
and $\sigma$ spin. $V_{\gamma i}$ are the hopping matrix element between
the Co 3$d$-orbitals and the ligand states. We also include the 3$d$ spin-orbit (s.o.)
in its first-order perturbative form. The inclusion of the TM s.o. is significant for processes
that involve changes in the total spin of the electrons in the 3$d$ shell. 

\indent
Given the large changes in reorganization
energy that can accompany these processes, we need to couple our model, which deals only with the
valence electronic sector, to a relevant reaction coordinate, such as bond length change. This is especially vital for biomolecules where
function is modulated by structural changes \cite{austin}. The simplest way to couple the TM molecule to the environment is 
via the terms in Eqn.(4): The term $g_{\gamma}xd_{\gamma\sigma}^{\dag}d_{\gamma\sigma}$ provides a (Holstein) linear coupling of 
the $d$-level charge occupancy of the TM atom and the reaction coordinate, the change in metal-ligand bond distance, $x$. At lowest order, the potential 
energy surfaces are harmonic and so we can add an energy term that is parabolic in $x$: $(1/2)Kx^{2}$. 
The linear coupling and the parabolic
energy term can both be obtained from a gradient expansion of the hybridization about the M-L bond length minimum \cite{SSH}:
\begin{eqnarray}
V(x) = V(x)\mid_{x=x_{0}} + \frac{\partial V}{\partial x}\mid_{x=x_{0}}(x-x_{0}) + \\ \nonumber
\frac{1}{2!}\frac{\partial^{2} V}{\partial x^{2}}\mid_{x=x_{0}}(x-x_{0})^{2} + \cdot\cdot\cdot\label{grad}
\end{eqnarray}
where $V(x)\mid_{x=x_{0}}=V_{\gamma\sigma}$, the linear coupling
$\frac{\partial V}{\partial x}\mid_{x=x_{0}}$ we estimate by taking the first derivative of the
appropriate Slater-Koster integral \cite{slater}, using a distance-parameterized expression \cite{harrison},
\begin{equation}
-g:=\frac{\partial V}{\partial x}\mid_{x=x_{0}} \simeq \frac{\partial V_{pd\sigma}}{\partial x}
\mid_{x=x_{0}} = \eta_{pd\sigma}\frac{-7}{2}\frac{\hbar^{2}r_{d}^{3/2}}{mx^{9/2}}\mid_{x=x_{0}}\label{g}
\end{equation}  
where $x_{0}$ is the equilibrium bond length. This expression should be valid for small displacements
away from $x_{0}$ and weak coupling between the TM $d$-electrons and changes in M-L bond length. 
From a simple harmonic oscillator model $K$ is related to the single-active
metal-ligand stretching mode $\omega$ by $\omega = \sqrt{K/M}$ where $M$ is the reduced mass which may be
approximated by the total mass of all six NH$_{3}$ ligands. To obtain $K$ we refer to experimental metal-ligand vibrational data \cite{richardson}.

\indent
We now show how we apply this model to a particular system, the self-exchange electron-transfer
in the cobalt hexaammines, $Co(NH_{3})_{6(aq)}^{2+/3+}$.

\section{Electron-Transfer in $Co(NH_{3})_{6(aq)}^{2+/3+}$}

\subsection{Experimental and Theoretical Review}

\indent
We would like to examine how effects arising from strong correlations manifest in the
general problem of electron transfer between solvated 1$^{st}$-row transition metal ions that are
mediated by a ligand bridge. In particular, we focus on the electron
self-exchange rate in the cobalt hexaammine system $Co(NH_{3})_{6(aq)}^{2+/3+}$, which is
given by the following reaction:
\begin{eqnarray}
Co(NH_{3})_{6(aq)}^{2+} + Co(NH_{3})_{6(aq)}^{3+} \stackrel{k_{et}}{\rightleftharpoons} \\ \nonumber
Co(NH_{3})_{6(aq)}^{3+} + Co(NH_{3})_{6(aq)}^{2+}
\end{eqnarray}
and the reasons why $k_{et}$ is two orders of magnitude larger than predicted.
Experimental measurements of the rate ($k_{et}\sim$10$^{-6}$-10$^{-5}$ M$^{-1}$s$^{-1}$) \cite{expts} deem the reaction adiabatic, i.e. where
the electron transmission coefficient
$\kappa\approx 1$, while previous theoretical efforts ($k_{et}\sim$10$^{-8}$ M$^{-1}$s$^{-1}$) yield $\kappa_{el} << 1$ \cite{buhks79}.
Electronic structure effects can be important for ET between small TM complexes in solution
(which we define to be the metal ion and the six coordinating ligands of the first
solvation shell) that are in contact with each other. In the case of the hexaammines,
the salient feature appears to be a "spin barrier", (first pointed out by Orgel \cite{orgel}).
During the course of the transfer process, each cobalt ion must change its spin by $\Delta S = 3/2$.
Strictly speaking, $k_{et}$ should be zero, but various schemes have been proposed to show
why the rate is, in fact, almost adiabatic.

\indent
The model for the electron-transfer process is usually treated as two cobalt ions (one Co(II) valence, the other Co(III))
that are each surrounded by their primary solvation shells which consist of six octahedrally-coordinated NH$_{3}$ groups.
The ET between the Co(NH$_{3}$)$_{6}^{2+/3+}$   
complexes is generally believed to be an outer-sphere process, so the ligands
act as a bridge that the tunnelling electron delocalizes over during the reaction.
The redox-active orbitals of the ammine ligands are the highest occupied molecular orbitals (HOMOs) of
the ammines which have $\sigma$ symmetry and hybridize strongly with the cobalt $e_g$ orbitals \cite{newton86}.
The strong $\sigma$-donating ability of the ammine ligand has been
confirmed by NMR studies which observe the kinetic stability
of this complex through the rate of water exchange of Co(NH$_{3}$)$_{6}^{2+/3+}$. This rate for
Co(NH$_{3}$)$_{6}^{2+/3+}$ is a factor of $\sim$10$^{6}$
slower than for Co(H$_{2}$O)$_{6}^{2+/3+}$ \cite{aqueouschemistry}.

The ET reaction can be reduced to a unimolecular process by consideration of the encounter complex, which is formed by the van der Waals contact
of the two primary coordination shells of the metal ions. 
This can be done in a variety of orientations \cite{newton03rev}. The two most commonly considered are the 
(1) apex-to-apex, which involves contact along the $C_{4}$ axis,
and is the most optimal bridge since the 1$s$ orbitals of the hydrogen atoms of the two ammines in close-contact show large mixing, and
(2) face-to-face, which has two faces perpendicular to the $C_{3}$ body-diagonal. The ammine contact is probably not
as good here but there are 6 ammines in contact instead of 2. For calculation of $k_{et}$ and $H_{DA}$, some averaging
over different orientations should be done to give a more realistic estimates of these quantities. We do not do that here,
only focusing on the most optimal geometry, apex-to-apex. Given this fixed geometry, we investigate electronic structure effects.
 
\indent
Buhks et al. \cite{buhks79} first noted that the groundstate-to-groundstate electron-transfer between the low-spin Co(III) ion (3$d^{6}$)($^{1}$A$_{1}$) 
and the high-spin Co(II) ion (3$d^{7}$)($^{4}$T$_{1}$) is spin-forbidden and the reaction must proceed
through the $S = 1/2$ $^{2}$E excited state for Co(II) or the $S = 1$ $^{3}$T$_{1}$
and $^{3}$T$_{2}$ of Co(III). They also showed that there is a non-zero s.o. coupling matrix element between
these ground and excited states, so the groundstates are not single-determinant but show small admixture to these excited states. Thermal
population of these excited configurations is an open issue \cite{newton1st,endreslabute,larsson86}. The vertical energy separation between $^{1}$A$_{1}$ and $^{3}$T$_{1}$
for Co(III) is very large ($\sim$14,000 cm$^{-1}$) and well-established by experiment. On the other hand, the separation between $^{4}$T$_{1}$
and $^{2}$E for Co(II) has been estimated from semi-empirical INDO/CI calculations to be 3000-9000 cm$^{-1}$ \cite{endreslabute}. 
While the existence of low-spin Co(III) has 
been experimentally verified, no such confirmation exists for high-spin Co(II). Thus, appreciable population of the $^{2}$E may yet be a 
viable possibility. Also, it should be noted that these values have been
estimated for the equilibrium configuration. The relevant values for the energy separation should be made at the transition state, where
they will be significantly reduced. This discussion is summarized in Fig.(1).    
\vspace{0.0cm}
\begin{figure}
\includegraphics[height=6.1cm,angle=0]{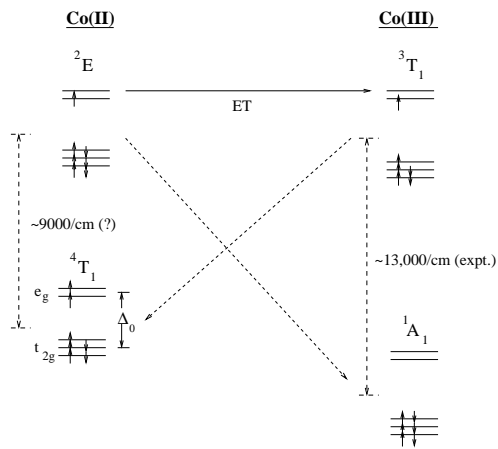}
\caption{Lowest energy configurations and excited states for Co$^{2+}$ and Co$^{3+}$ valences.
These are the relevant electronic states for the self-exchange electron transfer reaction
between cobalt atoms in Co(NH$_{3}$)$_{6}^{2+/3+}$. The question mark for the Co$^{2+}$ excited
state refers to an estimate by Buhks, {\it et al.} at the equlibrium configuration.}
\end{figure}

\indent
Other theoretical work, most notably by Newton \cite{newton86} and Sutin \cite{sutin1st},
who used {\it ab initio} self-consistent field, spin-restricted Hartree-Fock to calculate the
electronic coupling matrix element $H_{DA}$, assumed $^{2}$E as the Co(II) groundstate.
Newton obtained a value of 940 cm$^{-1}$ for the spin-allowed electron transfer. He then
used a reduction factor of 2 X 10$^{-2}$ to account for the 2$^{nd}$-order spin-orbit coupling,
obtaining a value of 20 cm$^{-1}$ for the $^{4}$T$_{1}$ to $^{1}$A$_{1}$ process which corresponds to
a $\kappa\simeq$ 4.0 X 10$^{-3}$. Furthermore, Newton considered different orientations
for the encounter complex, such as apex-to-apex, edge-to-edge,
and face-to-face, which will alter hybridization-matrix elements and differ in the number
of bridging ligands (2, 4, 6, respectively).

\indent
Newton, in a more recent paper \cite{newton1st}, calculated the relative energies of
$^{4}$T$_{1}$ and $^{2}$E electronic configurations of Co(II) and the $^{1}$A$_{1}$, $^{3}$T$_{1}$ of Co(III)
using unrestricted Hartree-Fock (UHF) and second-order M\o ller-Plesset (UMP2). He
recalculated $H_{DA}$ and $\kappa$ for the apex-to-apex geometry using these semi-empirical CI methods that take into account electron
correlations at the level of 2nd-order Brillouin-Wigner perturbation theory. For the groundstate
pathway, $^{4}$T$_{1}$/$^{1}$A$_{1}\rightarrow ^{1}$A$_{1}$/$^{4}$T$_{1}$
he obtains a values ($H_{DA}$ = 9.5 cm$^{-1}$, $\kappa$ = 1.2 X 10$^{-3}$)
which are optimal in the sense that the $3z^{2}-r^{2}$ - NH$_{3}$ $\sigma$-HOMO
hybridization can only decrease in other geometries. He estimates the effect
of orientational averaging by taking into account a multiplicative factor
0.1;  $\bar{\kappa_{el}}\sim$0.1($\kappa_{el}$)$_{apex-to-apex}$ so
$\bar{\kappa_{el}}\sim$10$^{-4}$, still falling 2 orders of magnitude short of
explaining the observed results.

\subsection{Many-Electron Basis Set}

\indent
We use the model Hamiltonian of Eqn.(1), which involves use of  
an {\it ansatz} for the groundstate wavefunction
with the coefficients to be determined variationally. The wavefunction is composed 
of the lowest-lying single-determinants in the many-electron Hilbert space, where the truncation
of the space is dictated by energetics. Our problem therefore reduces to diagonalizing the model Hamiltonian in a restricted
Hilbert space. We first considered the two nominal valences for each cobalt hexammine reactant. In
each case the most significant states are the high-spin and low-spin state,
the spin-orbit excited states, and states that couple by hole transport to the bridging NH$_{3}$ HOMO.
For Co(NH$_{3}$)$_{6}^{2+}$ we use the groundstate $^{4}$T$_{1}$, which is the three-fold
degenerate $hs$-3$d^{7}$. The spin-orbit state is, as stated above, $ls$-3$d^{7}$.
Each of the manifolds couples to respective hole-transfer states which are labeled
by $O_{h}$ irreps of the 3$d$ electron sector  $^{3}$A$_{2}$, $^{1}$A$_{1}$, $^{1}$E, $^{3}$T$_{1}$, and $^{3}$T$_{2}$
of the 3$d^{8}$ configuration on the cobalt atom. Due to suppression of $d^{9}$ occupancy by $U_{d}$,
we restrict the Hilbert space to the one-hole sector on the ligands. The states we retain as our basis gives a total of 65 states.

\indent
For Co(NH$_{3}$)$_{6}^{3+}$, we consider the following 3$d^{6}$ states: The experimentally
observed groundstate configuration $^{1}$A$_{1}$, and the states that couples via spin-orbit
coupling $^{3}$T$_{1}$, $^{3}$T$_{2}$. We would like to observe the suppression of the $d^{8}$
configuration, so we include many more hole-transfer states for this calculation.
For the 3$d^{7}\underline{L}$ states we have $^{2}$E and $^{4}$T$_{1}$. We also include   
the two-hole states (3$d^{8}\underline{L}^{2}$) $^{3}$A$_{2}$, $^{1}$A$_{1}$, $^{1}$E,
$^{3}$T$_{1}$, and $^{3}$T$_{2}$, where the hole-degeneracies are fully taken into account,   
giving us 376 states. For the case of both the 2+ and 3+ valences, we assume the truncation of
the basis set is justified by the suppression of all other $d^{n}$ configurations
by $U_{d}$ and all other spin states by the Hund's rule exchange interaction $J_{d}$.

\indent
The basis sets and the quantum weight distribution for Co(NH$_{3}$)$_{6}^{2+}$ and
Co(NH$_{3}$)$_{6}^{3+}$ are shown in Fig.(6). The next section
discusses how parameters were obtained for our model.

\subsection{Determination of model parameters}

\indent
Most of our model parameters are extracted from single complex calculations 
with the fully {\it ab initio} code SIESTA \cite{siesta} based on 
density functional theory (DFT). 
Before discussing the procedure in detail, we briefly want to introduce this code.
Nevertheless we want to stress that our way of determining the parameters is quite
general and independent of the code one uses as long as it has certain 
features such as a local atomic basis 
set and the projected density of states (PDOS).

In running SIESTA we used Troullier-Martins norm-conserving pseudo potentials \cite{TM} in the 
Kleinman-Bylander form \cite{KB}. For cobalt, we included spin-polarization 
and non-linear core corrections \cite{nlc} to account for a spin-dependent 
exchange splitting and correlation effects between core and valence electrons, 
respectively. Relativistic effects are included for the core electrons in the 
usual scalar-relativistic approximation (mass-velocity and Darwin terms)
and by averaging over spin-orbit coupling terms, while no spin-orbit coupling is 
included for the 4s and 3d valence electrons. Hence spin stays a good quantum number.
SIESTA uses a local basis set of pseudo atomic orbitals (PAO) of 
multiple $\zeta$-type. The first-$\zeta$ orbitals 
are produced by the method by Sankey and Niklewski \cite{sankey}, while the 
higher-$\zeta$ orbitals are obtained from the split valence method well known from quantum
chemistry. Polarization orbitals can also be included. 
We used a double-$\zeta$ basis set with polarization orbitals (DZP),
as well as the generalized gradient approximation (GGA) in the version by 
Perdew, Burke and Ernzerhof \cite{PBE} for the exchange-correlation energy functional. 
As for the single complex calculations the structures were optimized with the 
conjugate gradient method untill the forces on each atom were below a 
tolerance of $0.01 eV/\AA$.

\begin{figure}[tb]
\includegraphics[height=8.5cm,angle=-90]{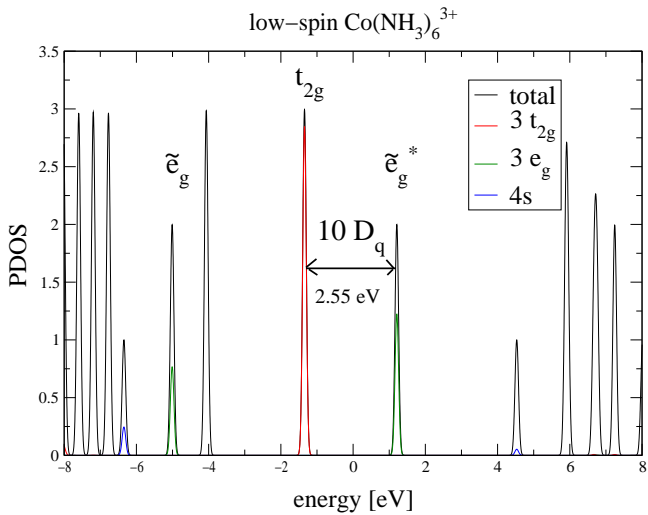}
\caption{\label{pdos_co3}Projected density-of-states (PDOS) of low-spin Co(III) ($d^6, S=0$) for spin-up and spin-down. The Fermi energy $\epsilon_F=-11.82 eV$ is set to zero. }
\end{figure}

\begin{figure}[tb]
\includegraphics[height=8.5cm,angle=-90]{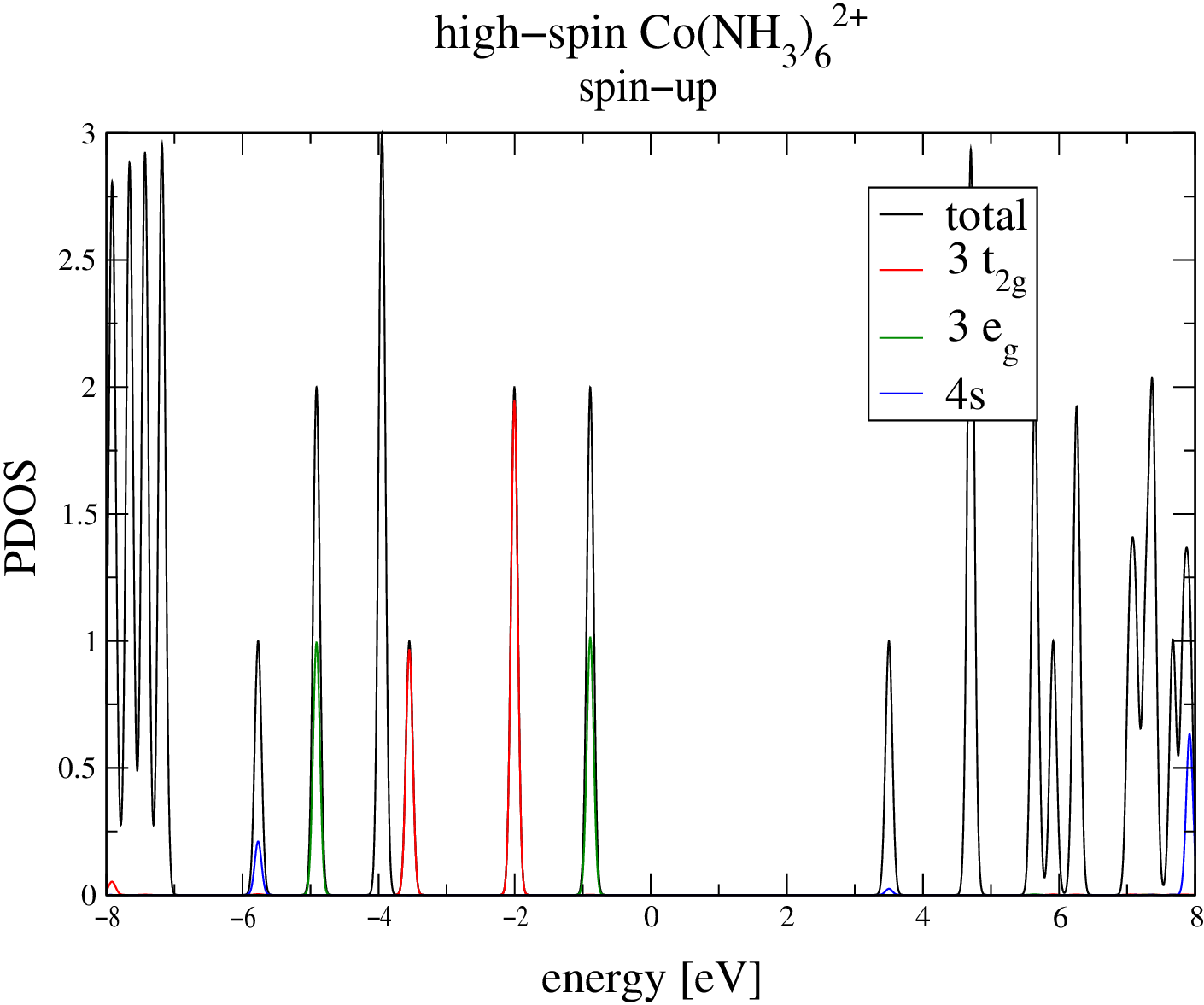}
\caption{\label{pdos_co2_up}PDOS of high-spin Co(II) ($d^7, S=3/2$). The spin-up
part is shown. The Fermi energy $\epsilon_F=-5.77 eV$ is set to zero.}
\end{figure}

\begin{figure}[tb]
\includegraphics[height=8.5cm,angle=-90]{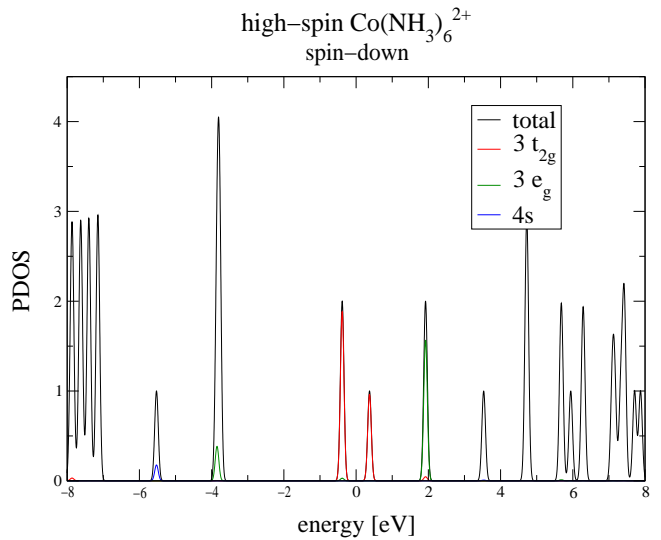}
\caption{\label{pdos_co2_down}PDOS of high-spin Co(II) ($d^7, S=3/2$). The spin-down part is shown. The Fermi energy $\epsilon_F=-8.17 eV$ is set to zero. }
\end{figure}

\begin{figure}[tb]
\includegraphics[height=8.0cm,angle=0]{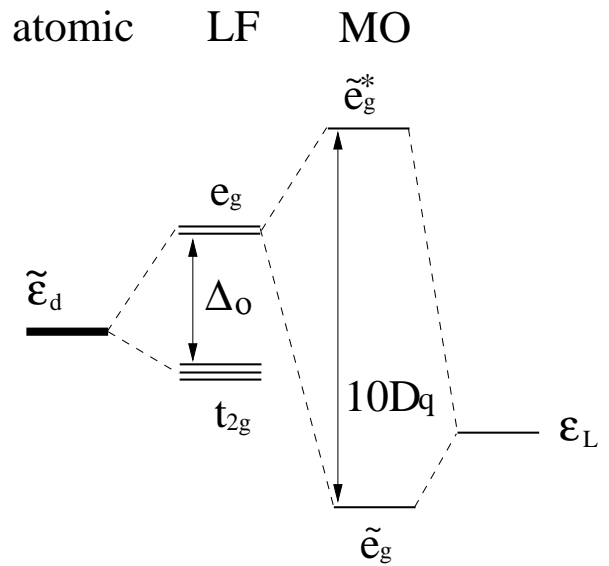}
\caption{\label{LF-MO}Ligand-field (LF) and molecular orbital (MO) pictures.}
\end{figure}

We first discuss the estimation of the following parameters, the d-d Coulomb repulsion $U_d$, 
Hund's coupling $J_d$, bare (undressed) d-orbital energy level 
$\epsilon_d$, the ligand energy level $\epsilon_L$, the ligand-field (purily 
due to electrostatics) $\Delta_o$, and $10D_q$ which contains contributions from both 
electrostatics {\it and} hybridization. In order to get a physical picture of the single complex
problem we show the PDOS (DOS projected on atomic 4s, 3 $t_{2g}$ and 3 $e_g$) of Co(III) in figure \ref{pdos_co3}
as well as Co(II, spin-up) in figure \ref{pdos_co2_up} and  Co(II, spin-down) in figure \ref{pdos_co2_down}.  
All states below the Fermi energy $\epsilon_F$ (set to zero) are occupied. 
One can immediately see the $10D_q$ energy gap between $t_{2g}$ and the anti-bonding $\tilde e_g^*$ orbitals
due to the approximately octrahedral symmetry. The $\pi$-symmetric $t_{2g}$ orbitals hardly hybridize with the 
$\sigma$-symmetric HOMO of $NH_3$
while the $\sigma$-symmetric $e_g$ orbitals form bonding
$\tilde e_g$ and anti-bonding $\tilde e_g^*$ molecular orbitals (MOs). For further clarification this 
is schematically illustrated in figure \ref{LF-MO}.
As for Co(II) the non-vanishing total spin (S=3/2) introduces spin-dependent MOs due to Hund's coupling,   
and the partly occupied $t_{2g}$ orbitals split probably due to rhombic distortions. Interestingly, the 
$e_g$ weights are of similar magnitude in the bonding $\tilde e_g$ and the antibonding $\tilde e_g^*$ MOs.
This indicates the near degeneracy of the $e_g$ and ligand energy level $\epsilon_d$ (HOMO of $NH_3$). 
In the case of Co(II, spin-up) [see Fig. \ref{pdos_co2_up}] they are exactly degenerate 
which provides the ligand energy level $\epsilon_L^{2+}=0.5(\tilde e_g^*+\tilde e_g)$ immidiately. 
The $t_{2g}$ and $10D_q$ energies of Co(II/III) can be read off the PDOS, too. In the case of
Co(II) we use the weighted average of the $t_{2g}$ peaks. The remaining parameters 
are estimated as follows.  

In order to get the ligand energy level $\epsilon_L^{3+}$ of Co(III)
we use a simple two-level approximation to describe the mixing of the Co-NH$_{3}$ levels.
In matrix form this is
\begin{eqnarray}
\left( \begin{array}{ll}
e_g &  V  \\
V   &  \epsilon_{L}
\end{array}\right)\label{matrix},
\end{eqnarray}
\indent
where $V$ is the hybridization between the Co $e_{g}$ levels and the ammine level $\epsilon_L$.
Setting the two eigenvalues of Eq. (\ref{matrix}) equal to $\tilde e_g$ and $\tilde e_g^*$
and also constraining the coefficients of the eigenvectors by the weights from the PDOS in 
figure \ref{pdos_co3}, we obtain $\epsilon_L^{3+}$ (as well as $V$ and $e_g$).

The Coulomb parameters were obtained from a crude mapping assumed between
the Kohn-Sham energies obtained from the GGA calculation and a mean-field calculation,
where we make use of the single-particle energies of the Co $t_{2g}$ levels
\begin{eqnarray}
&&\tilde\epsilon^{DFT}_{t2g,\sigma} \approx \tilde\epsilon^{HF}_{t2g,\sigma} = \nonumber \\
&& \epsilon_{d} + U_{d}(N_{d}-1)  
+ \frac{J_{d}}{5}(N_{d\sigma}-N_{d-\sigma}) - \frac{2}{5}\Delta_{0}\label{HF}
\end{eqnarray}
where $\epsilon_{d}$ is the bare energy (without any electron-electron interaction),
$N_{d\sigma}$ are the number of 3$d$ electrons with spin $\sigma$, $N_{d}$
is the total number of $d$-electrons, and $\Delta_{0}$ is the Madelung (point-charge)
contribution to the ligand field. We can write down three equations of this kind, one
for Co(III) (spin-independent) and two for Co(II) (one for spin-up and another one for 
spin-down), and can solve for the three unknowns $\epsilon_d$, $U_d$, and $J_d$. 
The required parameter $\Delta_o$ can be otained as follows.

Since we know the $t_{2g}$ energies from the PDOS we need to know the $e_g$ energies in order
to determine the electrostatic contribution $\Delta_o=e_g-t_{2g}$ for Co(II) and Co(III). These can be 
extracted in a straightforward way by looking sharply at figure \ref{LF-MO} which summarizes
all the involved quantities. From the center $\bar \epsilon=0.5(\tilde e_g+\tilde e_g^*)$ 
of the bonding $\tilde e_g$ and anti-bonding $\tilde e_g^*$ MOs 
and the knowledge of the ligand energy level $\epsilon_L$ we otain 
$e_g=\bar\epsilon+(\bar\epsilon-\epsilon_L)$. The resulting point charge contributions to the total
ligand-field $10D_q$ are small, $\Delta_o^{2+}=1.1 eV$ and $\Delta_o^{3+}=0.1 eV$. This is because
DFT seems to be biased towards forming rather extended MOs due to an improper treatment of the
on-site repulsion among d electrons leading to the well-known deficiency of over-binding \cite{martinNiO}.  
It does not seem to be surprising that $\Delta_o^{2+}>\Delta_o^{3+}$ since due to a $0.2 \AA$ 
shorter Co-N bond length in the case of Co(III) the hybridization is even larger than for Co(II).
Within DFT the $10D_q$ of Co(III) is mainly due to hybridization.

The ability of an electron to transfer from one cobalt ion to the other relies on the 
hybridization between the HOMO's on the two bridging ligands, which is estimated from the splitting of
the HOMO peaks of a two 2 ammine calculation. The ammines are in the apex-to-apex orientation.
The splitting of the HOMO corresponds to 2$t_{inter}$ where we define $t_{inter}$ to be the 
inter complex NH$_{3}$-NH$_{3}$ hopping integral. A similar procedure was applied to obtain the intra complex
NH$_{3}$-NH$_{3}$ hopping integral $t_{intra}$ between ammines of the same complex allowing 
holes to delocalize among the ligands.
The $Co-NH_3$ hybridization V was determined by computing the expectation values
$<\!x^2-y^2/z^2|H|HOMO\!>$ for Co(II) and Co(III) using the HOMO wavefunction 
from a single ammine and the Hamiltonian matrix of a single complex calculation. 
Contributions to the hybridization 
from direct overlap of non-orthogonal d and HOMO wavefunctions were neglected.

Finally, we determine the phonon-coupling constant g and the spring constant K as already outlined
in section \ref{section2}. According to Eq. (\ref{g}) g depends on the Slater-Koster parameter 
$\eta_{pd\sigma}$. It is determined by setting $V_{pd\sigma}=V$ from the DFT method. As for K 
we use the experimental Co-ligand stretching frequencies, $357 cm^{-1}$ for Co(II) and $494 cm^{-1}$ 
for Co(III) \cite{endicott}.     

The parameters are summarized in Table I. Therefore we can write down 
an effective Hamiltonian that includes only these frontier orbitals: 
the set of Co 3$d$ orbitals and the NH$_{3}$
HOMO states. We use the Hamiltonian of Eq.(1) as the effective Hamiltonian for each
of the single cobalt hexaammine complexes, the 2+ and 3+ nominal valences. The $d$-electron
parameters $\epsilon_{d}$, $U_{d}$, $J_{d}$, the ligand-field splitting 10$Dq$, etc.
all refer to the cobalt 3$d$ electrons, while the ligand label $L$ refers to the ammine HOMO.
We would like to solve for the groundstates of Co(NH$_{3}$)$_{6}^{2+/3+}$ within this
model.

\begin{table}
\caption{\label{tab:example} Parameters determined for hexaammine cobalt complexes
from SIESTA calculations. (1) and (2) refer to in-plane and axial ligands, respectively.
All parameters except $g_{\gamma}$(eV/$\AA$) and $K$(eV/$\AA^{2}$) are in units of eV.)}
\begin{ruledtabular}
\begin{tabular}{lll}
  Parameters & Co(II) & Co(III) \\
  $\epsilon_{d}$     &   -33.9    &  -33.9      \\
  $U_{d}$	     &     4.1    &    4.1      \\ 
  $J_{d}$	     &    -2.0    &   -2.0	\\
  $g_{\gamma}$	     &    -3.8    &   -5.2	\\
  $K$		     &    48.0    &   92.0	\\
  $\xi$              &     0.09	  &    0.09	\\
  10$Dq$             &     1.6	  &    2.6	\\
  $\epsilon_{H}$     &   -11.1	  &  -14.4	\\
  $V(x^{2}-y^{2}-L)$ &     1.9	  &    2.8	\\
  $V(z^{2}-L)$ (1)   &    -1.1	  &   -1.6	\\
  $V(z^{2}-L)$ (2)   &    -2.1    &   -3.2	\\
\end{tabular}
\end{ruledtabular}
\end{table}

\section{Results and discussion}
\subsection{Quantum weights of single complexes}

\indent
The model Hamiltonian we constructed is diagonalized on
the many-electron basis we have described using the set of parameters presented  
in Table I. We show the results in Fig.(6).

\vspace{0.0cm}
\begin{figure}[tb]
\includegraphics[height=6.0cm,angle=0]{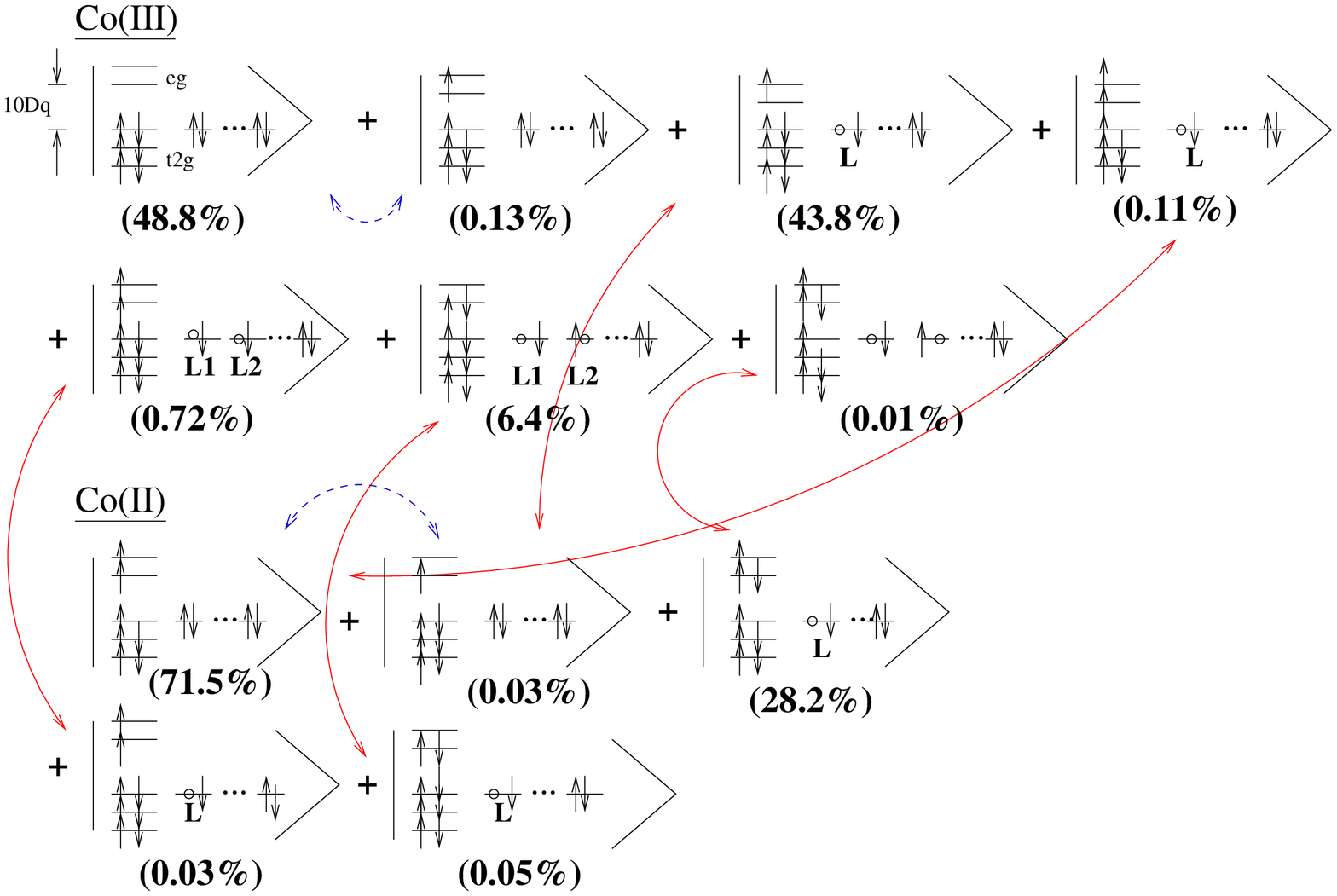}
\caption{Many-body groundstates of Co(NH$^{3}$)$_{6}^{3+}$ and Co(NH$^{3}$)$_{6}^{2+}$. The quantum weights
are the squared expansion coefficients of the groundstate eigenvector $\mid\alpha_{i}\mid^{2}$
on the restricted Hilbert space. The red arrows indicate
states that would couple through the bridge Hamiltonian. The blue arrows show coupling
through the spin-orbit term. There are 65 basis states for Co(NH$_{3}$)$_{6}^{2+}$ and 376
for Co(NH$_{3}$)$_{6}^{3+}$.}
\end{figure}

\indent
For Co(NH$_{3}$)$_{6}^{2+}$, there is 71.5\% in the $hs$-3$d^{7}$ ($^{4}$T$_{1}$) state and 28.1\% in the corresponding  
3$d^{8}\underline{L}^{2}$ where an electron has transferred from the NH$_{3}$ HOMO to the
$e_{g}$ shell of the Co atom. This weight is actually distributed with $\sim$5\% on
each of the six coordinated ammine ligands. This distribution of quantum weight is due to the NH$_{3}$-NH$_{3}$ hybridization
we included on our model. One may picture the hole as delocalized or {\it smeared} over the first solvation shell.
The fact that only a small fraction of the hole
resides on any one ligand has important implications for the charge-transfer process and we
will discuss this later. The remainder of the quantum weight resides in the $ls$-3$d^{7}$ ($^{2}$E),
which is small since it mixes into the groundstate by the weak spin-orbit coupling. The vertical
energy separation $E(^{2}E-^{4}T_{1})$ is 1.18 eV (9513 cm$^{-1}$), which is in good agreement with
{\it ab initio} calculations.

\indent
For Co(NH$^{3}$)$_{6}^{3+}$, the vertical energy separation $E(^{3}T_{1}-^{1}A_{1})$ is too small ($\sim$0.5 eV)
when compared with experiment ($\sim$1.7 eV) for the set of parameters. A parameter that we have
some freedom with is the ligand-field splitting 10$Dq$. The lack of constraint on 10$Dq$ emerges from
the model not taking proper account of all of the contributions to the ligand-field energy
splitting. 10$Dq$ is composed of the point-charge ligand-field $\Delta_{0}$ and a contribution
due to the hybridization of the cobalt and ligand orbitals. While the model does take into
account the most important hybridization between the $e_{g}$ and ammine HOMO, it {\it does not} actually
form bonding/anti-bonding pairs since the 3$d$ electron sector of our basis set is represented in
the atomic orbital basis, i.e. the $e_{g}$ levels are just the atomic orbitals, so in the current treatment
we cannot accurately predict the hybridization contribution. Therefore, we treat the ligand-field splitting
as a free-parameter in our model. While the 10$Dq$ value seems to work well for Co(NH$^{3}$)$_{6}^{2+}$ it does not reproduce
the correct vertical energy separation between low-spin and $S = 1$ 3$d^{6}$, so by increasing it by 1 eV
to 3.55 eV, we recover the vertical energy from experiment. The effect on the weights is to
shift $\sim$5$\%$ from $ls$-3$d^{7}$ to $ls$-3$d^{6}$, as would be expected.

\indent
We now need to determine if and how the effects of correlation between the $d$ electrons manifest
in this system. For Co(NH$^{3}$)$_{6}^{3+}$, the hole seems nearly fully delocalized between the first
coordination sphere of ammines and the cobalt 3$d$ shell with 51\% in the
$ls$-3$d^{6}$ ($^{1}$A$_{1}$)
configuration and 42\% in the $ls$-3$d^{7}\underline{L}$ states. This weight is divided up with
$\sim$7\% per state with a hole localized in one of the six NH$_{3}$ HOMO's. At first
glance, the physical picture appears to be very much like the screening cloud associated with the Kondo effect. In the case
of cerium materials, the hole left behind from an electron hopping from the conduction band
to the 4$f$ impurity delocalizes over the entire continuum of conduction states, dynamically screening
the local moment and leading to the formation of the singlet bound state.

\indent
We cannot say if the appreciable hole-weight delocalized over the ligands is due to mixed valence state or not.
Both the $ls$-3$d^{6}$($^{1}$A$_{1}$)
and $ls$-3$d^{7}$($^{2}$E) states possess filled diamagnetic $t_{2g}$
shells, so the Hund's rule exchange
contribution to the $d$-$d$ Coulomb interaction is equal in each
case. Since we only have
two configurations, we may shift the Fermi level to correspond with the
$ls$-3$d^{7}$, therefore we do not have to refer to $U_{d}$. From this perspective,
the delocalization of the hole between metal 3$d$ and the ammine orbital
looks simply like the consequence
of single-particle covalency. This interpretation is also consistent with
the composition of molecular
orbitals from our {\it ab initio} calculations which show nearly equal
contributions from metal
and ligand orbitals in the ($e_{g}$-$\sigma$) bonding/anti-bonding pair.

\indent
However, there does seem to be an analogy between the $d^{6}$,
$d^{7}\underline{L}$, and $d^{8}\underline{L}^{2}$
configurations of Co(NH$_{3}$)$_{6}^{3+}$ and the configurations $f^{0}$,
$f^{1}\underline{L}$, and $f^{2}\underline{L}^{2}$
of the mixed valence rare-earth compounds such as CeO$_{2}$ and metallic
$\alpha$-Ce, where mixed valence
has been established. In both the transition metal
and lanthanide cases these multiple configurations mix into the
groundstate because of their near-degeneracy.
In both cases there is also large suppression of the two ligand-hole states
due to the Hubbard-$U$ Coulomb repulsion
of electrons on the localized impurity levels. In the cerium materials 
the $f^{1}$ electron and the ligand hole $\underline{L}$ can form an exciton-like boundstate that
preserves the symmetry of the diamagnetic
$f^{0}$ state, i.e. a spin-singlet state with no orbital moment. In
Co(NH$_{3}$)$_{6}^{3+}$, we would have a similar energy
lowering associated with freezing out the 2-fold orbital degeneracy of the
$e_{g}$ shell and the spin-degree of
freedom, antiferromagnetically coupled to the hole delocalized over the
six NH$_{3}$ HOMO levels.

\indent
Such a state usually forms if the
singly-occupied configuration is slightly lower   
in energy than the empty level. This requirement is not satisfied in Co(NH$_{3}$)$_{6}^{3+}$
since the diamagnetic 3$d^{6}$
state is stabilized by 1.6 eV with respect to the $ls$-3$d^{7}$
configuration. However, the hybridization between
the $e_{g}$ and ammine HOMO's are large enough ($\sim$2-3 eV) to
mitigate this energy gap. In CeO$_{2}$
the partial occupancy of the 4$f$-level is observed to be $\sim$0.6, consistent
with the lattice volumes that are somewhat shorter than expected for
Ce$^{3+}$ compounds. In Co(NH$_{3}$)$_{6}^{3+}$, if
this type of partial occupancy exists ($\sim$0.5 from our model
calculation), experimentalists should
observe bond lengths that are somewhere intermediate between those
expected for Co(NH$_{3}$)$_{6}^{2+}$ and Co(NH$_{3}$)$_{6}^{3+}$.                                                                                     
\indent
For Co(NH$_{3}$)$_{6}^{2+}$, as stated above, the values for the single-particle Co 3$d$ and ammine HOMO
energies are such that there is almost complete resonance, and molecular
orbitals are constructed with $\sim$50\% of each type of orbital. We
observe in our calculated result the effect of the large $U_{d}$ in
suppressing the 3$d^{8}$ weight so the admixture is now on the order
of $\sim$70$\%$ of the hole is localized in the $d$ shell, while
$\sim$30$\%$ is delocalized over the ammine ligands. This suppression
of 3$d^{8}$ configurations is expected. The question is what consequence
do these effects have on the electron transfer matrix element $H_{DA}$.

\subsection{Construction of Born-Oppenheimer Potential Energy Surfaces and Estimate for $H_{DA}$}

\indent
We performed calculations of the groundstates for the single
Co(NH$_{3}$)$_{6}^{2+}$ and Co(NH$_{3}$)$_{6}^{3+}$ complexes. We would now like to use the basis sets
of both of these complexes to form an encounter complex that represents
the reactants in the electron-transfer process.

\indent
We connect the single 2+/3+ complexes in the apex-to-apex orientation
through a bridge composed of ammines from each molecule, which gives us the Hamiltonian
\begin{eqnarray}
H = h_{1} + h_{2} -t\sum_{\sigma}(c_{LA\sigma}^{\dag}c_{RA\sigma} + h.c.) 
\end{eqnarray}
Both h$_{1}$ and h$_{2}$ refer to the identical model Hamiltonian for each single complex given by Eqn.(1). 
The third term describes hole tunneling through the ligand bridge composed of the two ammines in
close contact. The creation operator $c_{LA\sigma}^{\dag}$ refers to the creation
of a $\sigma$-spin hole on the left NH$_{3}$ ligand (belonging to Co(NH$_{3}$)$_{6}^{3+}$) and $c_{RA\sigma}^{\dag}$
is the corresponding operator for the right ammine (belonging to Co(NH$_{3}$)$_{6}^{2+}$). The hopping integral $t$
can be estimated by noting that 2$t$ is approximately the energy splitting between
the bonding/anti-bonding orbitals formed
from the HOMO of each of the two bridging NH$_{3}$, oriented by close contact of
the hydrogen atoms of each ammine.
We must now solve for the groundstate
using the appropriate basis set. We construct this by taking a tensor product of our basis sets
for Co(NH$_{3}$)$_{6}^{2+}$ and Co(NH$_{3}$)$_{6}^{3+}$. The basis set is doubled to account for exchanging
Co(NH$_{3}$)$_{6}^{2+}$ and Co(NH$_{3}$)$_{6}^{3+}$ in space. There is then another multiplication by two in order
to treat the Hermitian matrix in order to use diagonalization routines suited for real,
symmetric matrices \cite{numerical}. So the basis set is then of rank
\begin{eqnarray}
N = 376 \times 65 \times 2 \times 2 = 97760.
\end{eqnarray}
Given that most routines scale as $\sim N^{3}$, where $N$ is the dimensionality
of the Hilbert space, conventional routines, even on very fast machines ($\sim$4 GHz
microprocessor) would make single-point calculations very time-consuming. Given that
we would like to sample the 2$D$ groundstate energy surface as a function of the breathing
coordinates for each complex, $x_{L}$ and $x_{R}$ at the resolution needed to accurately
determine the transition state, we need very efficient matrix diagonalizer. Plotting this 
potential energy surface is a computationally intensive problem.

The solution was to implement a L$\acute{a}$nczos matrix-diagonalization routine, which scales as order $N$ \cite{cullum}
since it involves only matrix-vector multiplication for sparse matrices.
The L$\acute{a}$nczos algorithm converts a large $N\times N$ matrix into a tridiagonal $m\times m$ matrix ($m\le M$).
It uses a three-term recurrence relationship which requires one trial groundstate vector to start the sequence. The 
eigenvectors tend to become parallel during iteration of the procedure. Since we are not interested
in them for this application, a costly ($\sim N^{3}$) reorthogonalization step does not have to be implemented.

The plot in Fig.(9) is a two-dimensional contour plot representing the
Born-Oppenheimer potential energy surface we
have plotted solving the 97760$^{2}$
eigenvalue problem 1600 times.

\vspace{0.0cm}
\begin{figure}[tb]
\includegraphics[height=8.0cm,angle=0]{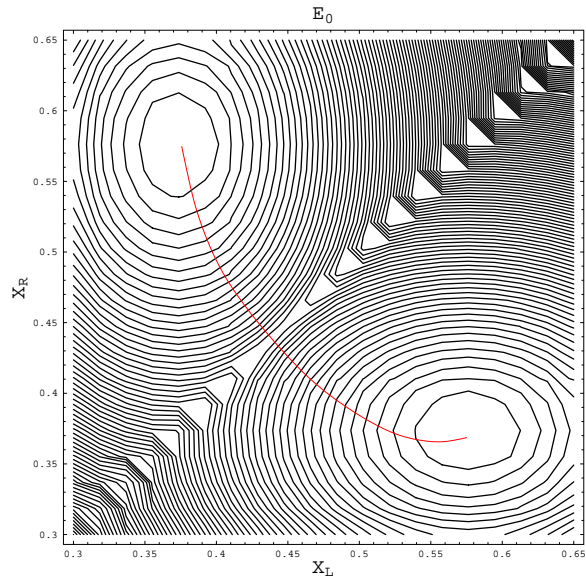}
\caption{The Born-Oppenheimer potential energy surface for the Co(NH$_{3}$)$_{6}^{2+/3+}$
self-exchange electron transfer reaction. Only the inner-sphere contribution is shown. The 
groundstate energy was plotted on a 40$\times$40 grid using a L$\acute{a}$nczos diagonlization routine to diagonalize 
the 97760$^{2}$ eigenvalue problem. The contour level spacing is chosen to be 0.03 eV/level.
$x_{L}$ and $x_{R}$ are the reaction coordinates corresponding to the totally symmetric 
metal-ligand stretching mode. The transition state, defined as the saddle point along $x$=
$x_{L}$=$x_{R}$ is at $x$=0.44$\AA$. The difference in the Co-N bond length between Co(II)
and Co(III) is 0.2$\AA$.}
\end{figure}

The oval shape of the minima are consistent with
a larger force constant (narrower potential) for Co(NH$_{3}$)$_{6}^{3+}$. The horizontal and vertical
distances between the minima correspond to $\sim$0.2$\AA$ which agrees well with the
experimentally measured difference in the metal-ligand distance between the 2+ and 3+ forms of the
hexaammine cobalt complex. The donor surface {\bf D} for the ET corresponds to
the encounter complex Co(NH$_{3}$)$_{5}^{3+}$(NH$_{3}$)-(H$_{3}$N)Co(NH$_{3}$)$_{5}^{2+}$,
which is given by the parabola in the
upper left corner. The acceptor surface {\bf A} is that of
Co(NH$_{3}$)$_{5}^{2+}$(NH$_{3}$)-(H$_{3}$N)Co(NH$_{3}$)$_{5}^{3+}$
and is the lower right parabola in Fig.(7). The transition state is the cusp feature at $x$=0.44$\AA$
and corresponds to a narrowly-avoided crossing between the {\bf D} and {\bf A} diabatic
surfaces. The electronic coupling opens a gap equal to 2$\mid H_{DA}\mid$ between {\bf D} and {\bf A}
at the crossing point, where $H_{DA}$ is the electronic coupling factor. 

\vspace{0.0cm}
\begin{figure}[tb]
\includegraphics[height=6.0cm,angle=0]{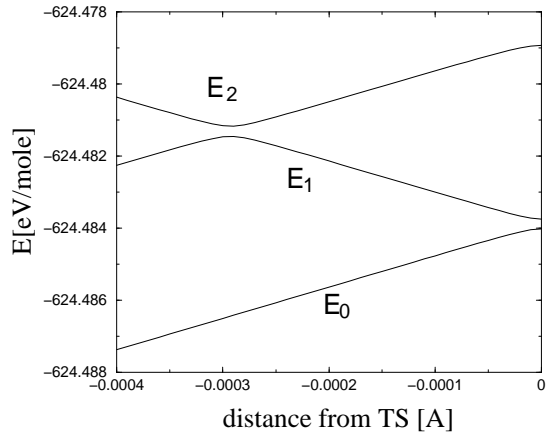}
\caption{Cross-section of PES from Fig.(7) along the reaction coordinate $d$=$x_{L}$-$x_{R}$ close to
the TS. The energy splittings (2$\mid H_{DA}\mid$) between the groundstate $E_{0}$ and first excited
state $E_{1}$ at the TS (0$\AA$) is about 2 cm$^{-1}$. The splitting between $E_{1}$ and $E_{2}$ at
$d$=-2.8$\times$10$^{-4}\AA$ has a similar value.}
\end{figure}

\indent
The general method of extracting $H_{DA}$ is as follows: We find the TS by plotting the energy along
the diagonal $x$ in Fig.(7) with high resolution. The TS is the saddlepoint at $x$=0.44$\AA$. We then
read off the difference between the first excited and ground states. $H_{DA}$ is half of this difference.

\indent
On the other hand, one can exploit that $H_{DA}$ is also a groundstate property. Instead of calculating the exicted
state, one can obtain it from calculating the energy difference between the groundstate adiabatic surface 
evalulated at the TS and the avoided crossing between the diabatic {\bf D} and {\bf A} surfaces that results
when the bridge hopping integral $t$ is turned off. 

\indent
In Fig.(10) we plot a cross-section of PES from Fig.(9) along the reaction coordinate 
$d$ = x$_{L}$-x$_{R}$ in the region of the TS. The energy splittings (2$\mid H_{DA}\mid$)
between the groundstate $E_{0}$ and first excited state $E_{1}$ at the TS (0$\AA$) is
$\simeq$2 cm$^{-1}$. The splitting between $E_{1}$ and $E_{2}$ at $d$=-2.8$\times$10$^{-4}\AA$
has a similar value. The splittings between $E_{0}$ and $E_{1}$ and also $E_{1}$ and $E_{2}$
originate from lifting of the degeneracy of the components of the $^{4}T_{1}$ irreducible
representation of Co(II) by the spin-orbit coupling.

\indent
Our result ($\sim$2 cm$^{-1}$) is smaller than Newton's calculation of $H_{DA}$, which
gave 9.5 cm$^{-1}$ from {\it ab initio} calculations. We have essentially reproduced
this calculation from our simpler model. What the reduction of $\mid H_{DA}\mid$ implies
is that the correlation effects we have observed in the single complexes of
cobalt hexaammine do not seem to enhance the ET rate. The reason is that
we still rely on states that couple through the spin-orbit coupling, most noticeably
$ls$-3$d^{7}$($^{2}$E) for the Co(NH$_{3}$)$_{6}^{2+}$ complex. The quantum weight of this state
in the groundstate is not increased by the correlation-induced suppresion
of 3$d^{8}$ occupancy. In fact, this suppression tends to reduce the magnitude of 
$H_{DA}$. The arrows in Fig.(2) indicate that the 3$d^{8}$
states of both Co(NH$_{3}$)$_{6}^{2+}$ and Co(NH$_{3}$)$_{6}^{3+}$ couple through the bridge term
of $H$. The weight on these states is small because of the large $U_{d}$
value.

\section{Conclusions}

\indent
We have introduced a variant of the  Anderson Impurity Model that
is appropriate for describing the valence electronic structure of TM molecules.
We have coupled this model to a metal-ligand stretching mode. We have applied this
model to investigate the effect of strong correlations 
among the cobalt 3$d$ electrons on the self-exchange electron transfer rate in the cobalt
hexammines Co(NH$_{3}$)$_{6}^{2+/3+}$.

\indent
We have constrained the model parameters by using the results of
{\it ab initio} density functional theory calculations and experimental stretching frequencies.
The single-particle hybridizations $V_{dL}$ were
extracted directly and the Coulomb parameters
($\epsilon_{d}$,$U_{d}$,$J_{d}$) were obtained   
from mapping to a mean-field solution of our model.
Using these parameters, we found the groundstates
of both 2+ and 3+ nominal valences.
For the Co(NH$_{3}$)$_{6}^{3+}$ state we find strong configurational
admixture between the 3$d^{6}$ and 3$d^{7}\underline{L}$
states. The hole $\underline{L}$ is fully delocalized
between the cobalt atom and the first coordination sphere
of ammines. While this resembles the screening cloud
and partial occupancy of the impurity level seen in
mixed valent compounds like CeO$_{2}$, the large 3$d^{7}\underline{L}$ weight could
simply be a purely single-particle effect since the delocalization
is consistent with the equal metal and ligand orbital composition
of molecular orbitals near the Fermi level from the DFT calculation.
For Co(NH$_{3}$)$_{6}^{2+}$ there is an obvious signature of the effect
of large $U_{d}$, with a suppression of $\sim$20$\%$ of the quantum
weight on 3$d^{8}\underline{L}^{2}$ states.

\indent
We couple the complexes together by a bridging Hamiltonian
to form the encounter complex and generate a basis set
by using the tensor product of the single hexaammine basis states.
This gives a very large eigenvalue problem ($N\simeq$10$^{5}$) that
is solved using L$\acute{a}$nczos methods. The groundstate energy surface
is plotted, resolving the two minima of the electron transfer process.
We are able to extract the electronic coupling $H_{DA}$ as the energy
lowering at the transition state, relative to the diabatic surfaces.
What we found is that configurational admixture does not enhance
$H_{DA}$ and we get a result $\sim$4 cm$^{-1}$, that is within a
factor of 5 of Newton's original calculation.

\indent
It appears that the disparity of two orders of magnitude
in the self-exchange electron transfer rate of Co(NH$_{3}$)$_{6}^{2+/3+}$
cannot be explained on the basis of the electronic
structure of a single complex. One aspect that has been completely ignored in
this calculation are solvation effects beyond the first coordination   
shell. While one would expect, in the gas phase, that the single, symmetric breathing coordinate
would be an excellent approximation to the breathing coordinate, solvent effects
could significantly change the heigth of the barrier by adding anharmonicity to the
potential wells. This could contribute to the adiabatic nature of the reaction.

\indent
Another possible source of enhancement for the ET rate could be through 
{\it dynamical} effects. The dynamical prefactor depends on the details
of how the system crosses and recrosses the thermal barrier. The presence
of the low-lying spin-orbit states that are separated from the groundstate
by very small ($\sim$1 cm$^{-1}$) energies could give the vibronic wavepacket
multiple channels for reaching the other minimum. This is a possibility that
can be explored with our model, which provides fast "on-the-fly" sampling of the
correct electronic structure. Excited state calculations of the bimetallic TM active site
involving open-shell states, coupled by the TM atom spin-orbit interaction, would be very 
difficult to do with state-of-the-art quantum chemistry codes.

\indent
Finally, it is possible that the relevant Born-Oppenheimer surfaces at the transition state are
different from those at equilibrium, which we have explored elsewhere within DFT
theory\cite{endreslabute}.  In particular, an intermediate spin surface appears to be thermally
accessible and favorable at the transition state.  In that paper we actually found a rate two orders of
magnitude {\it faster} than experiment.  It is interesting to note that within our calculation and
Newton's UMP2 calculationi\cite{newton1st}
the electron correlation effects actually {\it reduce} the transition rate.
Within our variational state treatment the reason is clear:  Coulombic blocking effects eliminate a
class of determinants which can facilitate charge transfer.  It will be interesting to explore the
effects of correlations upon the intermediate spin manifold of states.

\indent
The method proposed in this paper could be applied to other TM systems such
as the active sites in metalloproteins. More elaborate ways of coupling the electronic
sector to the protein structure will have to be devised but the efficiency and simplicity
of our method might be very useful in identifying transition state configurations.

{\bf Acknowledgements}  We have benefitted from useful discussions with M. Newton. This work was 
supported by the United States Department of Energy, Office of Basic Energy Sciences, Division of
Materials Research. M.X.L. has been supported by a UC Davis NSF NEAT-IGERT
graduate fellowship and the Laboratory Directed Research and Development program at
Los Alamos National Laboratory.

\end{document}